# Parrondo's Paradox:  new results and new ideas

**Abhijit Kar Gupta\***
Associate Professor
Department of Physics, Panskura Banamali College, Panskura, Dist. East Midnapore, Pin: 721 152, WB, India
e-mail: kg.abhi@gmail.com

**Sourabh Banerjee**
Assistant Professor
Department of Physics, Chandernagore Govt. College, Chandernagore, Dist. Hooghly, Pin: 712 136, WB, India
e-mail: bandosou@gmail.com

## Abstract:

*Parrondo's paradox is about a paradoxical game and gambling. Imagine two kinds of probability dependent games A and B, mediated by coin tossing. Each of them, when played separately and repeatedly, results in losing which means the average wealth keeps on decreasing. The paradox appears when the games are played together in random or periodic sequences; the combination of two losing games results into a winning game! While the counterintuitive result is interesting in itself, the model can very well be thought of a discretized version of Brownian flashing ratchets which are employed to understand noise induced order. There are a plenty of examples from physics to biology and in social sciences where the stochastic thermal fluctuations or other kinds actually help achieving positive movements. It is in this context, the Brownian ratchets and the kind of prototype games may be explored in detail.*

*In our study, we examine various random combinations of losing probabilistic games in order to understand how and how far the losing combinations result in winning. Further, we devise an alternative model to study the similar paradox and examine the idea of paradox in it. The work is mostly done by computer simulations. Analytical calculations to support this work, is under progress.*

## Introducing the Original Parrondo's Paradox Game:

Parrondo's paradox is about probabilistic games or gambling. In this case, the games are played with biased coins. There is no doubt that we are destined to lose (losing our capital in a game in the long run) while playing (repeatedly) with an unfair or a biased coin. Let us imagine two differently designed losing games; they are so when played separately and repeatedly. Now, if we play the games together, can we win? The answer is surprisingly yes! This is counterintuitive to think and is known as Parrondo's paradox as that was first devised by a physicist, Juan M.R. Parrondo[1].

In the following, we introduce the original Parrondo's Paradox scenario (Figure 1). Consider, two simple coin tossing games A and B. Also, as it happens in gambling, capital of the player is linked with it. If we win, our capital would increase by 1 unit and loosing makes our capital to decrease by 1 unit. Game A is played with a coin which is slightly biased towards losing. The probability of winning is $p = 1/2 - \varepsilon$ and the probability of losing is $1 - p = 1/2 + \varepsilon$, where $\varepsilon = 0.005$, a very small number which acts as a biasing parameter. Game B is little more complex. If the capital is not modulo 3 (that is not divisible by 3), then we are to play with a biased coin towards winning: with the probability of winning being $p_1 = 3/4 - \varepsilon$ and so the probability of losing is $1/4 + \varepsilon$. Let us refer to it as a '**good coin**'. When the capital is a multiple of 3, we have to play with a heavily biased coin towards losing and in this case the probability of winning is $p_2 = 1/10 - \varepsilon$ and so the probability of losing, $9/10 + \varepsilon$. Let us refer to this as a '**bad coin**'.

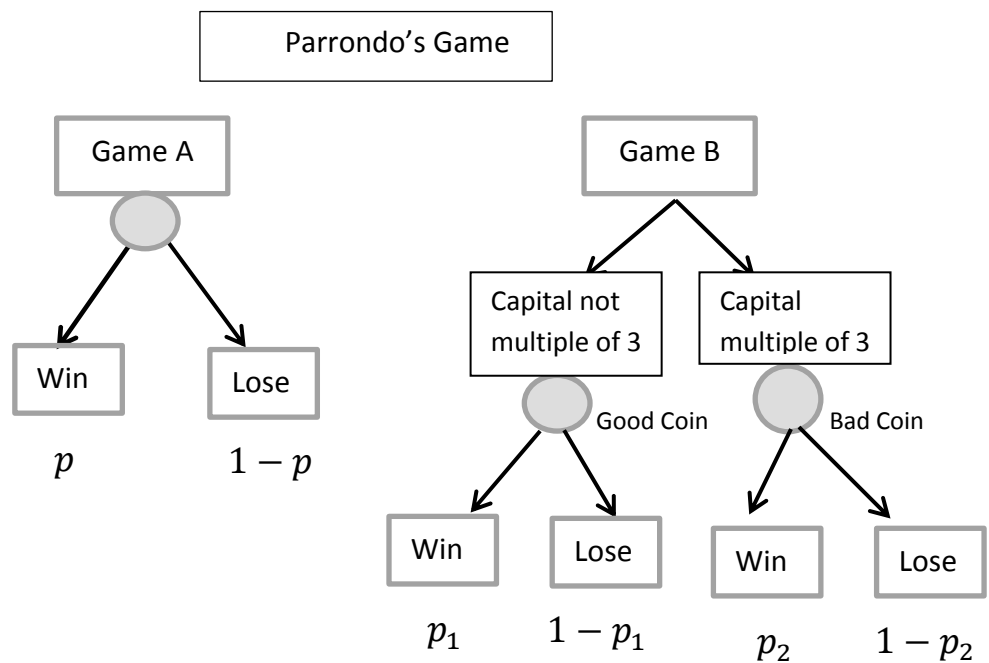

**Figure 1:** Parrondo's Paradox Game: pictorial demonstration. In the original game, and for our consideration here in this work, $p = 0.5 - 0.005$, $p_1 = 3/4 - 0.005$ and $p_2 = 1/10 - 0.005$.

Game A and game B are losing games when played separately. Losing (winning) means the mean capital keeps on decreasing (increasing) with the number of turns of the games played. For game A, it is obvious to think. But for game B, one needs to go through some heuristic arguments based on Markov process to arrive at the conclusion. This can also be easily checked by numerical computer simulations which we have done.

In a beautiful recent review[2], Parrondo *et al.* described the mechanisms in great detail and clarity with illustrations. In this, they dealt with some mathematical aspects and outlined many scenarios and numerical results that originated from such a paradoxical game and its variations.

It is now definite, by design, that the two games A and B are losing games which means that the resulting capital goes on decreasing with the number of times when any of them is played. However, instead of playing the two losing games separately, if we alternate the games like AABBAABB....or in some other sequence or even in a random fashion, it is seen that the resulting capital goes on increasing with the number of turns they are played. The following figure [Figure 2] illustrates the claim where, for different situations, the average capital is plotted with the number of times the games played.

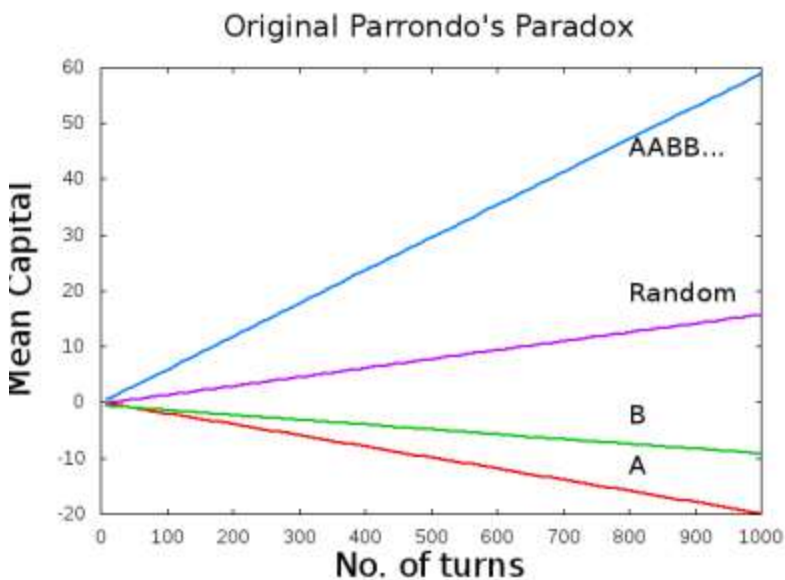

**Figure 2:** The demonstration of the original capital dependent game. The mean capitals against number of turns are plotted for four different cases. If the game A or B is played separately, it is clearly seen that the mean capital is decreasing monotonically and for the combined games (AABBAABB sequence or with random A and B), that is increasing monotonically.

The paradox is, at the first thought, very puzzling. But how is this possible? Game A is always a losing game. But game B is played with a bad coin and a good coin. It is sure that the number of times the bad coin is played can be controlled through the capital. Since the association of game A makes it winning, the effect must be that the number of times the bad coin is used must have been reduced when game A is played along with. This has been found to be true as checked by us numerically.  The effect of playing game A is to reorganize the capital in such a way that the capital turns less likely to be a multiple of 3. So the connection (between two games) and control is through the capital. This game was referred [in ref. 2] as capital dependent game. Incidentally, afterwards many other variations of Parrondo's game[3] and detailed numerical studies[4] appeared in literature, for example history dependent game where instead of capital the connection between two games is taken through the history of losing or winning[5].

Even though Parrondo's Paradox is a fancy simple model to lead us to investigate the counterintuitive situation, it opens up a new way to analyze and understand the functioning of Brownian ratchets and *flashing Brownian ratchets* in particular[6].  Brownian ratchets are connected with Brownian motors[7] which are some asymmetric potential dependent phenomena where directed move in presence of noise or fluctuations is possible. It is important to understand how tiny little protein based molecular motors move in the cell environment with intracellular thermal noise as well as under macroscopic environmental noise. There are plenty of applications and the interest lies in modeling in biology, physics, economics and other social sciences. A flashing Brownian ratchet is where the asymmetric potential is switched on and off in some fashion.

A simple flashing Brownian ratchet [in Figure 3] consists of a series of asymmetric potential on a line. If the particles are trapped in one minimum, they can not move due to asymmetric potential walls. In the absence of it, particles can diffuse freely on either sides, right or left. In this present scenario, in the absence of asymmetric potential, there is a force towards left with the help of a slanting potential. But when the ratchet potential is on again, many of the particles are forced to move towards right than on the left even with the presence of the small applied force towards left. Particles can move on the left when the ratchet potential is switched off or on. But it can be demonstrated that when the ratchet potential is made off and on randomly or periodically, a net rightward motion is possible even in the presence of the force toward left. This is called flashing ratchet and the two states (off and on of Ratchet potential) can be thought of as two games A and B with negative outcome.

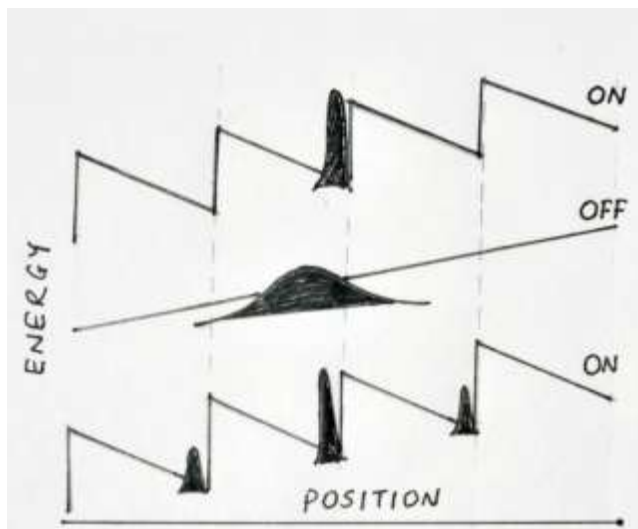

**Figure 3:** Demonstration of Flashing ratchet

Noise or fluctuation is a considerable nightmare in the fields of physics, biology or electrical engineering. But noise induced movement or stochastic resonance has lately gained importance in such fields. If we consider fluctuation is a probability then we can think of probabilistic games or gambling in order to understand the interplay of noises or fluctuations in the systems. Parrondo's Paradox is thus an approach where we can identify the roles of negativities that can bring out a positive outcome.

## Capital dependent Game: Random combinations of A and B

We have studied the Parrondo's paradox game, as outlined above, in more detail and tried to figure out how far the combination of two losing games A and B can give rise to winning situation. In this combined game, where the sole connection between games A and B is through the capital, the outcome of the game (win or loss) would depend on the relative appearance of game A with respect to game B. With some periodic combinations of A and B (for example, AABBAABB…), we have checked the resulting paradoxical behaviour, that is the resulting game is winning which means the mean capital of a player increases with the number of turns played. However, our idea has been to examine the combined game where one plays A and B in random order.

The questions to ask will the resulting game be always winning with any proportion of A or B played? To look for an answer, we simulated the combined games where game A is played with probability $p$ and game B with $(1 - p)$ in random order. For each value of $p[0,1]$, the simulation is done to check the resulting mean capital for reasonably long time steps to make sure of the trend, *i.e.*, if the combined game is winning or losing. But interestingly, it is observed that with any randomness, the combined game is not always winning. For a range of values of $p$ only, the game appears to be winning and for higher or lower values than that, the game is losing that means the paradox does not occur. The numerical result is displayed below. The understanding is that too much or too little presence of game A influences the capital in such a way that the triggering of the good coin in game B turns less and less. Game A changes the value of capital towards or away from the multiple of 3 which the game B cannot do alone. The mechanism in B changes the capital toward multiple of 3 more likely and that is the bad coin is played more often when game B is played only. These pathological scenarios are checked with detailed numerical investigations [Figure 4].

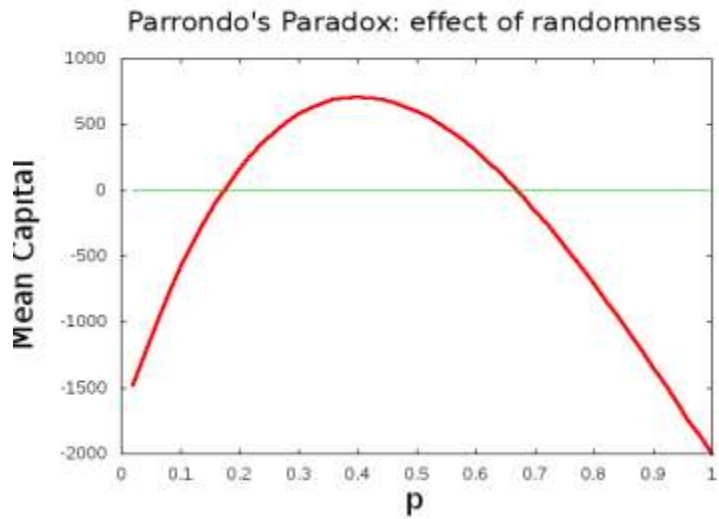

**Figure 4:** Variation of mean capital against the probability ($p$) of appearance of game A.

## A new History dependent Game:

It may be apparent that in the above described capital dependent game, the connection between two games A and B is the capital. The condition applied over capital (multiple of 3 or not) is what responsible for switching from one game to other and hence the effect of net outcome. If the similar switching can be done by some other means, presumably the similar paradox might happen. One such attempt has been done through the history of the games played. In Parrondo's works[5], such a history dependent game was devised where game A was kept intact and the game B to be played with 4 different biased coins depending on winning or losing in the previous two turns. However, in our new history dependent game that we devise, we obtain the paradox while playing the game B with two coins only.

In the new history dependent game (as described in the adjacent table), we keep the game A the same [Win with $0.5 - \epsilon$ and loss with $0.5 + \epsilon$] . The game B is played with two coins: Good coin and Bad coin, as before, with different probabilities. With Good Coin, the winning probability is $0.9 - \epsilon$ and with Bad Coin, the winning probability is $0.4 - \epsilon$. Clearly, the good coin is biased towards winning and the bad coin is biased towards losing. When to play with good coin or bad coin, will depend on the *history* of the previous two outcomes which is given in the following table.

| Step $t-2$ | Step $t-1$ | Coin used in Game B |
|---|---|---|
| Lose | Lose | Good coin [Win with $p = 0.9 - \epsilon$] |
| Lose | Win | Bad Coin [$p = 0.4 - \epsilon$] |
| Win | Lose | Bad Coin |
| Win | Win | Bad Coin |

**Table 1:** The New History Dependent Game. The possibilities with the two coins used.

Now, as before, like that of capital dependent game, when the games A and B are played separately, they are losing games in the sense that the average capital decreases with the number of turns. However, with a combination of them, for example, AABAAB… (where game A is played consecutively twice and then game B and so on…), the result is winning. A computer simulation is done and this is demonstrated in the following figure 5. [For mean capital, numerical results are obtained with 10,000 configurations in each case.]

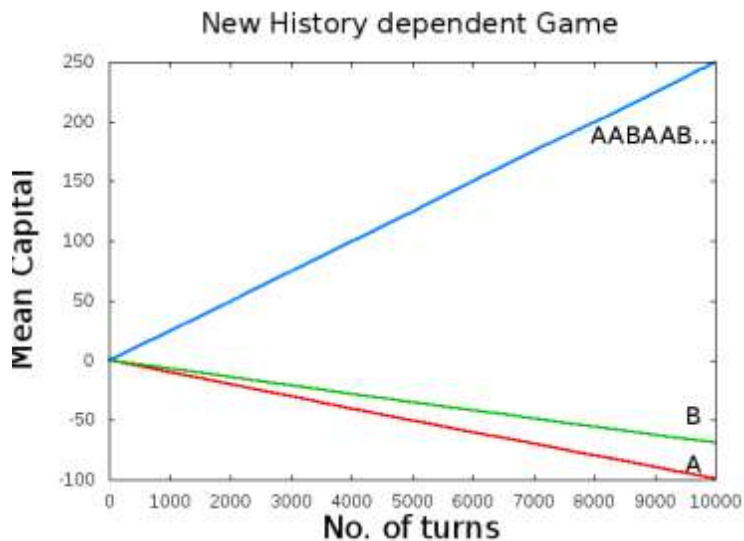

**Figure 5:** The evolution of mean capital against the number of turns when the games are individually played (A or B) and the case for a periodic combination of game A and game B.

The essence of Parrondo's paradox is that the combination of two losing games can give rise to a winning game. We have checked this with the original capital dependent game and then discovered that not all combinations can produce a winning game. For random combinations of A and B, we observed that the winning or losing of the combined game would depend on the probability of the game A (or B) appears. We checked the similar situation in this case too and the computer simulation shows a similar trend [Figure 6]. The mean capital (obtained after a certain number of turns) is first negative with lower probability, then positive for some intermediate values of probability ($p$), and finally it turns out negative again. This means, the combined game is losing for lower and higher probabilities and winning for a certain range of probability of appearance of game A.

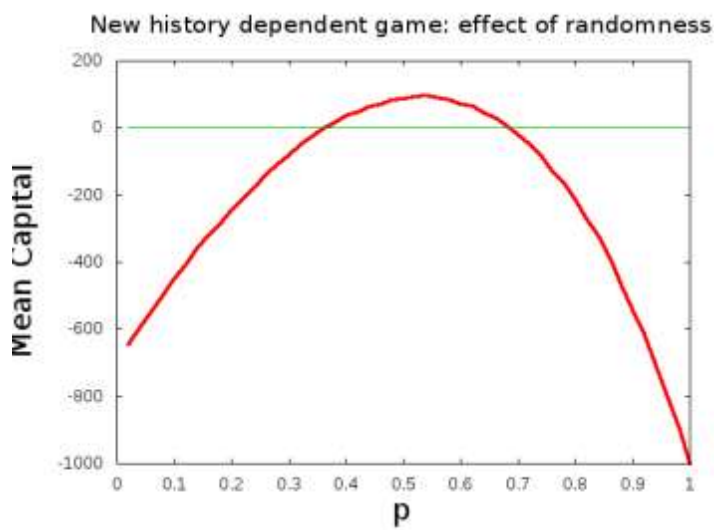

**Figure 6:** The variation of mean capital against the probability ($p$) of the game A that appears in the New History Dependent Game.

Further studies on the Parrondo's paradox games of several variations are done numerically to check various aspects of the games and the appearance of apparent paradox. One interesting and paradoxical result is obtained with capital dependent games, when the games are played by a group of players rather than by one individual player played so far. In this case, the decision of playing either game A or game B at some turn is dependent on optimization process[3].

## Conclusions:

We have studied original Parrondo's paradox where two losing probabilistic games (A and B) are combined to make a winning game. The two games can be combined in some ordered sequence or they can be selected randomly. However, our study reveals that the random selection of a game (A or B) plays an important role; the combined game can be winning or losing depending on the strength of randomness i.e., the probability ($p$) of appearance of a game (say, A). Two variations of Parrondo's paradox game have been studied: capital dependent and history dependent ones. We have devised a new history dependent game, a simpler version that the usual history dependent game studied so far. In this too, we have examined the role of relative appearance of a particular component game with respect to the other and we obtain the similar behaviour of winning or losing with respect to the strength of randomness. As a whole, our interest has been to study how and how far the paradox of winning from two losing games stands.